# The Disconnect Between Quantum Mechanics and Gravity

Daniel M. Greenberger
*Dept. of Physics, CCNY, New York, NY, 10031,USA.*
*greenbgr@sci.ccny.cuny.edu*

Abstract:  There is a serious disconnect between quantum theory and gravity.  It occurs at the level of the very foundations of quantum theory, and is far deeper than just the matter of trying to quantize a non-linear theory.  We shall examine some of the physical reasons for this disconnect and show how it manifests itself at the beginning, at the level of the equivalence principle.

The equivalence principle was the key insight that led Einstein to the general theory of relativity.  In this paper we assume that it is one of the truly basic laws of nature.  It defines locally what one means by an inertial frame, and frees one from having to consider the problem that always arises in Newtonian mechanics, namely "how does one know that one is in an inertial frame?"  If you are in free-fall in an external gravitational field, you are in the best approximation we have to an inertial frame.

One generally distinguishes between two forms of the equivalence principle, the weak form and the strong form.  First we shall consider the weak form and ask how it fares in quantum theory.  Then we will consider the strong form.

## The Classical Equivalence Principle

Before examining the equivalence principle in quantum theory, let us look at the classical form.  The weak principle describes how a point particle behaves in an external gravitational field, and the strong principle relates the motion to that of a particle in an accelerated reference frame.

Our own preference for describing the weak equivalence principle is the following.  If you place a point particle in a non-gravitational external field, a minor miracle occurs. It turns out that you can describe the behavior of the particle by introducing only one parameter, its mass. This one parameter is really all you need to describe its behavior. Nature could have been much more complicated, but luckily it chose not to be.

However in a gravitational external field, a major miracle occurs. One doesn't need any parameters at all. The motion is completely determined by the environment around the particle. In this case the external field completely determines the geometry around the particle, and the particle merely responds to this. The particle moves along a geodesic in this environment, and its trajectory is completely described by its position and velocity, so that all point particles will have the same behavior. So we expect that the position, **r**(t), and the velocity, **v**(t), in a given environment will be independent of the mass. The energy, $E = mv^2/2 + m\varphi$, should be then be linear in the mass. As a special case, when the field slowly disappears, we have a free particle moving in free space. In either case, one should not need a mass to describe the motion of the particle, and it should drop out of the formulation of the motion. (This is non-trivial even for a free particle. See the companion paper.)

The strong equivalence principle states that the motion of a particle in a gravitational field is equivalent to that of a particle at rest in an accelerating coordinate system. So the motion in an accelerated system is equivalent to being in free fall in the gravitational field, and again, the mass will drop out. So if you can solve the much simpler motion of a particle in an accelerated frame, you should know how the particle behaves in a gravitational field.

## The Weak Equivalence Principle in Quantum Theory

So, how does the weak equivalence principle fare in quantum theory? Well, if you put a particle into an external gravitational field, you immediately see a deep conflict. The problem is that according to quantum theory, the quantization principle is formulated in terms of momentum, rather than velocity, so that

$$\oint p\, dx = nh. \tag{1}$$

where p is the momentum. But the momentum has the particle mass built into it, p = mv. For example, if one had a free particle and one were constructing a minimum wave packet, then

$$\Delta x \Delta p \sim \hbar, \quad \Delta x \sim \hbar/\Delta p \sim \hbar/m\Delta v. \tag{2}$$

So the velocity spread depends on m. One can then determine the mass of the particle by measuring the velocity spread, which violates the spirit of the equivalence principle.

The same problem persists in a gravitational field. For example look at a gravitational Bohr atom, where one replaces the Coulomb potential by a gravitational one. If one has a small mass m orbiting a much larger one M in a 1/r potential, one can solve the Schroedinger equation by merely replacing the coupling constant $e^2$ in $V_C = e^2/r$ by GMm to get $V_G = GMm/r$, where G is the gravitational constant. If M is very large, it will act as an external potential, and classically, the mass of the light particle will drop out of all trajectory information. But if one solves the Schroedinger equation, one can find the gravitational Bohr radius in an S state, as

$$r_{n,Coul.} = \frac{\hbar^2 n^2}{me^2} \rightarrow r_{n,Grav.} = \frac{\hbar^2 n^2}{GMm^2},$$

$$E_{n,Coul.} = -\frac{e^2}{2r_n} = \frac{me^4}{2\hbar^2 n^2} \rightarrow E_{n,Grav.} = \frac{(GM)^2 m^3}{2\hbar^2 n^2}. \tag{3}$$

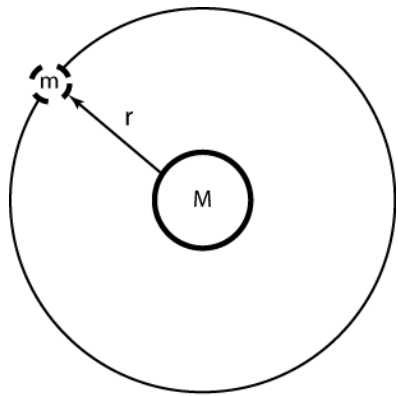

Fig. (1)

Fig. (1). The Gravitational Bohr Atom

We see here that the mass does not drop out of $r_n$, and the energy is not linear in m. So one could determine the mass of the lighter particle by measuring the radius, which totally violates the spirit of the classical equivalence principle. The problem, as we have noted, is that the mass is dynamically built into the system— even for a free particle.

One reason for this comes from eq. (1), which shows that the mass is built directly into quantum theory. Eq. (1) implies that

$$\oint v\, dx = nh/m. \tag{4}$$

In this equation, both the v and the x should classically be independent of the mass, but instead, the mass enters through quantization. There is also a deeper reason why equivalence fails, which we will discuss later.

## The Strong Equivalence Principle in Quantum Theory

The strong equivalence principle fares better in quantum theory, although there are still problems. To implement the strong equivalence principle one must move to an accelerated frame. For this the Galilean transformation is not enough as that boosts the system to a frame moving with constant velocity *v*. However, one may enhance the Galilean transformation and use what is usually called the Extended Galilean Transformation, which takes the system into a frame that is accelerating. However every point in space in the frame is accelerating at the same rate, so it is a rigid coordinate system. This is not a relativistic transformation, since distant points are always synchronized to the same speed and acceleration, so they must be able to communicate with each other instantaneously.

The extended Galilean transformation takes the system from (x,t) to (x',t'), where

$$x' = x + \xi(t), \quad t' = t;$$
$$\frac{\partial}{\partial x} \rightarrow \frac{\partial x'}{\partial x}\frac{\partial}{\partial x'} + \frac{\partial t'}{\partial x}\frac{\partial}{\partial t'} = \frac{\partial}{\partial x'},$$
$$\frac{\partial}{\partial t} \rightarrow \frac{\partial t'}{\partial t}\frac{\partial}{\partial t'} + \frac{\partial x'}{\partial t}\frac{\partial}{\partial x'} = \frac{\partial}{\partial t'} + \dot{\xi}(t)\frac{\partial}{\partial x'}.$$
(5)

The wave function *ψ(x,t)* becomes *ψ'(x',t')*, so the Schroedinger equation becomes

$$-\frac{\hbar^2}{2m}\frac{\partial^2 \psi'}{\partial x'^2} = i\hbar\left(\frac{\partial \psi'}{\partial t'} + \dot{\xi}\frac{\partial \psi'}{\partial x'}\right),$$
(6)

where now there is a momentum-dependent term on the right side. However, we can get rid of the $\partial/\partial x'$ term by introducing a phase into the wave function. Write

$$\psi'(x',t') = e^{if(x',t')}\varphi(x',t').$$
(7)

and then the derivatives in eq. (6) become

$$\frac{\partial}{\partial x'}e^{if}\varphi = (i\frac{\partial f}{\partial x'}\varphi + \frac{\partial \varphi}{\partial x'})e^{if},$$
$$\frac{\partial^2}{\partial x'^2}e^{if}\varphi = \left(-\left(\frac{\partial f}{\partial x'}\right)^2\varphi + 2i\frac{\partial f}{\partial x'}\frac{\partial \varphi}{\partial x'} + i\frac{\partial^2 f}{\partial x'^2}\varphi + \frac{\partial^2 \varphi}{\partial x'^2}\right)e^{if},$$
$$\frac{\partial}{\partial t}e^{if}\varphi = (i\frac{\partial f}{\partial t}\varphi + \frac{\partial \varphi}{\partial t})e^{if},$$
(8)

where we have replaced *t'* by *t*.

Then the Schroedinger equation becomes

$$-\frac{\hbar^2}{2m}\left(\varphi'' - f'^2\varphi + if''\varphi + 2if'\varphi'\right) = i\hbar\left[\left(\dot{\varphi} + i\dot{f}\varphi\right) + \dot{\xi}\left(if'\varphi + \varphi'\right)\right], \quad (9)$$

where we have replaced $\frac{\partial f}{\partial x'}$ by $f'$ and $\frac{\partial f}{\partial t}$ by $\dot{f}$, etc.

We have now to remove the terms in $\varphi'$, which we can do by equating its coefficients on both sides of eq. (9). The result is

$$\frac{\partial f(x',t)}{\partial x'} = -\frac{m}{\hbar}\dot{\xi}. \quad (10)$$

We can integrate this to obtain

$$f(x',t) = \frac{m}{\hbar}\left(-\dot{\xi}x' + g(t)\right). \quad (11)$$

Here, $g(t)$ is an arbitrary function to be determined.

We determine $g(t)$ so as to eliminate all the extra terms in eq. (9) that are of the form of an explicit time dependence, $u(t)\varphi$. This yields

$$\dot{g} = \frac{1}{2}\dot{\xi}^2, \quad g = \frac{1}{2}\int^t \dot{\xi}^2 dt. \quad (12)$$

So the final result is the Schroedinger equation in the accelerated system, for $\varphi(x',t)$,

$$-\frac{\hbar^2}{2m}\frac{\partial^2\varphi}{\partial x'^2} - m\ddot{\xi}x'\varphi = i\hbar\frac{\partial\varphi}{\partial t},$$

$$\psi(x',t) = e^{if(x',t)}\varphi(x',t), \quad f = \frac{m}{\hbar}\left(-\dot{\xi}x' + \frac{1}{2}\int\dot{\xi}^2\,dt\right). \quad (13)$$

This shows that in the accelerated frame, there is a gravitational potential, due to the acceleration, $-\ddot{\xi}$, which is exactly what one expects from the strong equivalence principle. However note that the mass still enters due to the phase factor, which will cause mass-dependent diffraction effects.

## The Connection of the Phase Factor to Proper Time

The phase that occurs in quantum theory is just a measure of proper time. It takes the form $e^{-imc^2\tau/\hbar}$. This reduces to the form $e^{i\int L\,dt/\hbar}$, which is the Feynman path integral form. For example, for a free particle,

$$e^{i(pr-Et)/\hbar} = e^{i(mv\gamma vt - m\gamma t)/\hbar} = e^{im\gamma t(v^2-1)} = e^{-im\tau}. \tag{14}$$

where we have assumed that $r = vt$, along the particle trajectory. In the non-relativistic limit this reduces to the form of the phase factor in eq. (13). To see this more clearly, look at the Lorentz transformation,

$$\begin{aligned} dt' &= \gamma(dt - v\,dx), \quad dx' = \gamma(dx - v\,dt), \\ dt' &= \gamma\,dt - v(dx' + \gamma v\,dt) = \gamma(1-v^2)dt - v\,dx' \\ &\xrightarrow{NR} dt - \tfrac{1}{2}v^2 dt - v\,dx' = dt + \delta\tau. \end{aligned} \tag{15}$$

Here we have used the fact that the phase in the Galilean transformation is expressed in terms of x' rather than x. Also, in our case, $v \to -\dot{\xi}$.
This yields

$$\begin{aligned} \delta\tau &= \dot{\xi}\,dx' - \tfrac{1}{2}\dot{\xi}^2 dt, \\ e^{-im(\tau+\delta\tau)/\hbar} &= e^{-im(\tau - \dot{\xi}\,dx' + \tfrac{1}{2}\dot{\xi}^2 dt)/\hbar}, \end{aligned} \tag{16}$$

where $\left(-\dfrac{m}{\hbar}\delta\tau\right)$ is just the phase factor in eq. (13). By the way, this also shows that the proper time leaves a residue in non-relativistic physics, so that the effects of the twin paradox also show up there, as phase factors. Thus, the mysterious phase factor that shows up in the Galilean transformation is there to take into account the non-relativistic residue of the proper time, in the form of the twin-paradox, even though classical physics does not recognize the concept!

## The Classical Limit of Equivalence

The deeper problem we referred to earlier concerns the classical limit of the equivalence principle, which is different from the classical limit for non-gravitational forces. The question then arises that if the mass shows up quantum mechanically, how does it disappear in the classical limit? This happens by virtue of a special scaling process that allows the mass to cancel out in this limit, but to show up for low-lying quantum states. Consider eq. (3), which shows how the mass enters into trajectory parameters where we would expect it to cancel out. Then in the classical limit, if you have two systems of different mass, $m_1$, and $m_2 = K m_1$, such that they have the same $r(t)$ and $v(t)$, it follows that if particle $m_1$ is in the state labelled by $n_1$, then particle $m_2$ must be in the state labelled by $n_2 = K n_1$.

$$\oint v_1 dr = \frac{n_1 h}{m_1}, \quad \oint v_2 dr = \frac{n_2 h}{m_2} = \frac{K n_1 h}{K m_1} = \oint v_1 dr. \tag{17}$$

Then the ratio of the two masses will drop out. So in the classical limit, it is not the masses that drop out, but their ratios. (The actual value of $m$ does not matter because in a wave packet one can choose an appropriate average value of $n$ such as to give the packet the correct classical values of $r$ and $v$.)

But this scaling can only happen in the classical limit, where the levels are very close together and linearly spaced. To see how this works, for two nearby levels,

$$\oint p_n dr = nh, \quad \oint p_{n+\ell} dr = (n+\ell)h =$$
$$= \oint (p_n + \Delta E \frac{\partial p}{\partial E}\bigg|_n) dr = nh + \Delta E \oint \frac{dr}{v_n}, \tag{18}$$
$$\Delta E T_n = \Delta E \frac{2\pi}{\omega_n} = \ell h, \quad \Delta E = \ell \hbar \omega_n.$$

Here we have used the fact that in the classical limit $\frac{\partial E}{\partial p} = v,$ from Hamilton's equations. We also assume that $\ell \ll n.$ For lower quantum numbers this scaling breaks down. One can see this scaling in phase space in Fig. (2).

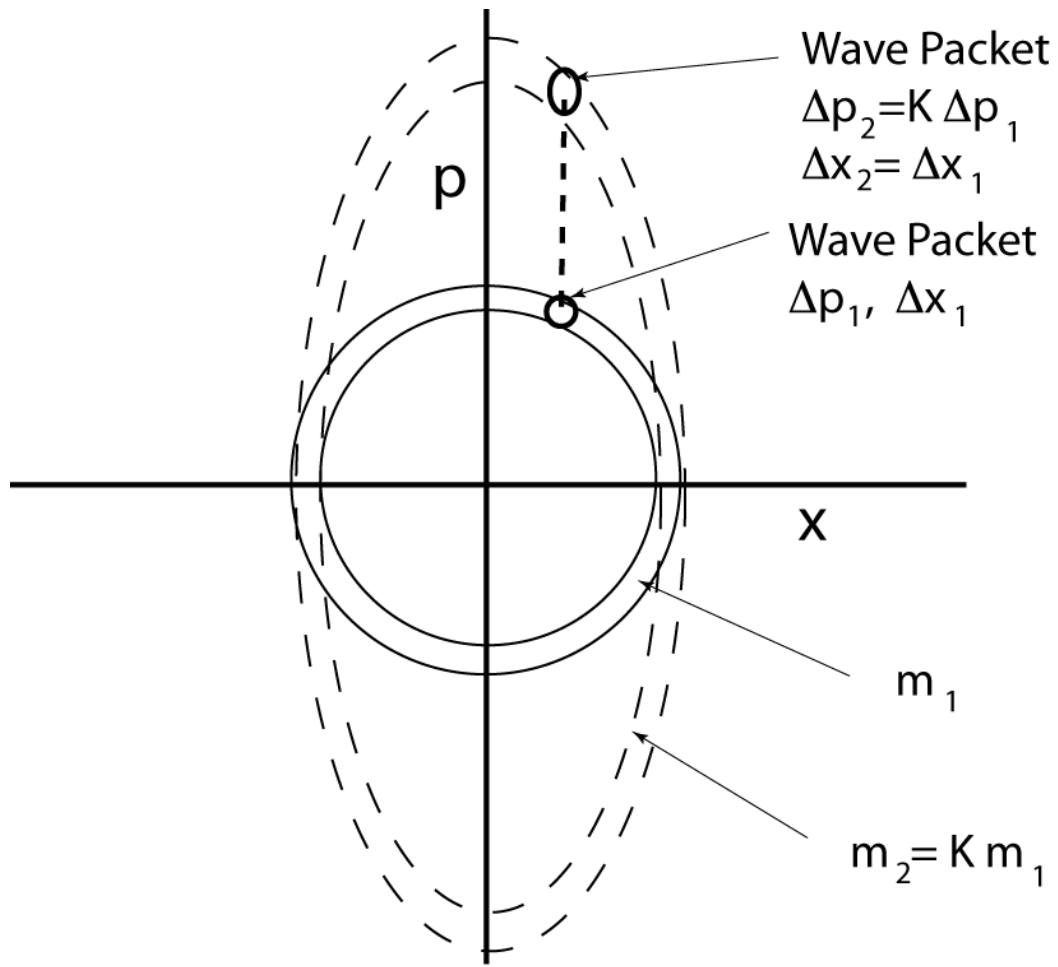

Fig. (2). <u>Classical Scaling of p in phase space.</u>

The scaling takes place because $p = mv$, and so $p$ scales with $m$, as does the energy. The situation is different for non-gravitational forces, where the momentum is important, and there is no equivalence principle. There one has

$$r(t) = \sum c_\ell e^{i\ell\omega t}, \quad c_\ell = \frac{1}{T}\int dt\, r(t)\, e^{-i\ell\omega t}. \tag{19}$$

The way this is approached is different for non-gravitational and gravitational forces, where

$$\begin{aligned} \text{Non-grav:} \quad & r(t) = \sum c_\ell e^{i\ell\omega t}, \quad \langle \psi_{n+\ell} | r | \psi_n \rangle \to c_\ell, \\ \text{Grav:} \quad & r(t;Km) = \sum c_\ell e^{i\ell\omega t}, \quad \langle \psi_{Kn+K\ell} | r | \psi_{Kn} \rangle \to c_\ell. \end{aligned} \tag{20}$$

So the correspondence limit is different in gravitational and non-gravitational fields, in order to accomodate the equivalence principle.

In a gravitational problem, quantum mechanically the mass enters differently for each potential, and there will be a separate mass-dependent unit of length for each one. For example in the Bohr problem we considered, it was $r_0 = \hbar^2/GMm^2$. For higher, semi-classical states, the ratio of masses drops out, as shown. But in the limit of low quantum numbers, there is a different $r_0(m)$ for each problem. This is the deeper problem I referred to earlier. There is no mass-independent fundamental length built into the theory, to serve as support for the equivalence principle to hold for low quantum numbers.

If quantum theory ever breaks down, this is where I would expect it to happen. I believe that the equivalence principle is a very basic principle of nature, and that it would be respected in very strong gravitational fields, which should then yield an alternate quantization principle that would be based on a fundamental length that does not exist in quantum theory. Then I would expect in the Bohr-Sommerfeld limit in a strong gravitational field, something like

$$\oint v\, dr = nc\lambda_0. \tag{21}$$

Here, $\lambda_0$ would be a real fundamental length in the theory. In weak gravitational fields, I would expect Planck's constant to exert itself. I might add that I don't know whether $\lambda_0$ would be very small or very large.

The natural question that occurs is why should $\lambda_0$ not be the Planck length,

$$\lambda_0 = \left(\frac{G\hbar}{c^3}\right)^{1/2} ? \tag{22}$$

The Planck length comes to about $10^{-34}$ cm, and it is mass-independent. But it is compiled of a combination of known constants, and assumes that no new physics occurs until it is reached. It is the place where one would expect quantum theory and gravity to interact, where space-time would dissolve into a quantum foam of some sort. But historically, such extrapolations have always been wrong, because new physics has always intervened. And we are a long, long way from $10^{-34}$ cm, much, much further than classical physicists were from atomic dimensions, where all sorts of unexpected results occured.

A similar situation occurred before quantum theory was invented, when people expected classical physics to break down at "the classical radius of the electron", which was obtained by setting $E = mc^2 = e^2/R$, where the coulomb energy of the electron reached its rest energy. A reasonable expectation, which gave $R = e^2/mc^2 \sim 10^{-13}$ cm. But before they reached this length, they encountered pair creation, and the Compton wavelength, $\lambda_C = h/mc \sim 10^{-11}$ cm. So such extrapolations are always dangerous. In that situation it was the advent of quantum theory. In this case, I tell my students that the extrapolation is like someone asking me where Prof. X is, and I reply that he should be at the North Pole. When they inquire why, I answer that I last saw him a month ago, and he was walking north, and I have no evidence that he has changed his direction since then, and I shouldn't make any speculative assumptions. So by now he should be at the north pole.

12That is conservative extrapolation to the point of lunacy, and I think that any extrapolations using the Planck length are no better.

Another reason is that the numbers we know are clearly missing something. For example, look at the Gravitational Bohr atom, from eq. (3). If it is used to estimate the Bohr radius of two neutrons gravitating around each other, one gets about $10^{27}$ cm, about the size of the universe. This not only makes no sense, it leaves large conceptual problems as to what a gravitational atom would mean.

A further argument along the same lines, is to try to estimate "nature's fundamental speed, $c^*$," assuming we knew all about gravity and quantum theory, but not electrodynamics. A reasonable estimate would be

$$c^* = \frac{Gm^2}{\hbar} \sim 10^{-30} \, cm/\sec, \tag{23}$$

where $m$ is the mass of the neutron. This insane estimate is closer to the speed of darkness than to that of light, and it is indicative of the fact that something is missing from our theories. Dirac once observed that all our dimensionless fundamental constants differ by powers of $10^{20}$, and we see that eq. (23) is off by about $10^{40}$. So I don't place any value in using the Planck length to predict new phenomena, and take $\lambda_0$ to be completely unknown. However I would expect it to lead to an era of completely new physics, where quantum theory and gravity would become compatible.